\documentclass[%
 reprint,
 amsmath,amssymb,
 aps,
 prl,
]{revtex4-1}

\usepackage{graphicx}
\usepackage{dcolumn}
\usepackage{bm}
\usepackage{fancyhdr}
\usepackage{extramarks}
\usepackage{amsmath}
\usepackage{amsthm}
\usepackage{amsfonts}
\usepackage{tikz}
\usepackage[plain]{algorithm}
\usepackage{algpseudocode}
\usepackage{braket}
\usepackage{multirow}
\usepackage{color}

\newcommand{\Yb}{\ensuremath{^{171}\mathrm{Yb}^+~}}

\begin{document}


\title{Realization of two-dimensional crystal of ions in a monolithic Paul trap}

\author{Ye Wang$^{1,2}$}
 \email{ye.wang2@duke.edu}
\author{Mu Qiao$^{1}$}
\author{Zhengyang Cai$^{1}$}
\author{Kuan Zhang$^{3,1}$}
\author{Naijun Jin$^{1}$}
\author{Pengfei Wang$^{1}$}
\author{Wentao Chen$^{1}$}
\author{Chunyang Luan$^{1}$}
\author{Haiyan Wang$^{1}$}
\author{Yipu Song$^{1}$}
\author{Dahyun Yum$^{1,4}$}
\author{Kihwan Kim$^{1}$}%
 \email{kimkihwan@mail.tsinghua.edu.cn }
\affiliation{%
$^{1}$ Center for Quantum Information, Institute for Interdisciplinary Information Sciences, Tsinghua University, Beijing 100084, P. R. China\\
$^{2}$Fitzpatrick Institute for Photonics, Electrical and Computer Engineering Department, Duke University, Durham, North Carolina 27708, USA\\
$^{3}$MOE Key Laboratory of Fundamental Physical Quantities Measurements,\\Hubei Key Laboratory of Gravitation and Quantum Physics, PGMF and School of Physics,\\Huazhong University of Science and Technology, Wuhan 430074, China\\
$^{4}$Department of physics and astronomy Seoul National University, South Korea 08826
}


\date{\today}

\begin{abstract}

We present a simple Paul trap that stably accommodates up to a couple of dozens \ensuremath{^{171}\mathrm{Yb}^+~} ions in a stationary two-dimensional lattice. The trap is constructed on a single plate of gold-plated laser-machined alumina and can produce a pancake-like pseudo-potential that makes ions form a self-assembly two-dimensional crystal which locates on the plane composed of axial and one of the transverse axes with around 5 $\mu$m spacing. We use Raman laser beams to coherently manipulate these ion-qubits where the net propagation direction is perpendicular to the plane of the crystal and micromotion. We perform the coherent operations and study the spectrum of vibrational modes through globally addressed Raman laser-beams on a dozen of ions in the two-dimensional crystal. We measure the amplitude of micro-motion by comparing the strengths of carrier and micro-motion sideband transitions with three ions, where the micro-motion amplitude is similar to that of a single ion. The spacings of ions are small enough for large coupling strengths, which is a favorable condition for two-dimensional quantum simulation.


\end{abstract}

\maketitle



Two-dimensional crystal of ions can be an attractive and natural platform to scale up the number of ion-qubits in a single trap and to explore many-body quantum models in two-dimension \cite{britton2012engineered,bohnet2016quantum,chiaverini2008laserless,schmied2009optimal,clark2009two,clark2011cryogenic,sterling2014fabrication,mielenz2016arrays,bermudez2011frustrated,bermudez2012quantum,nath2015hexagonal,yoshimura2015creation,richerme2016two,wang2015quantum,jain2018quantum,goodwin2016resolved}. Recently, a fully-connected quantum computer has been realized with up to 5-20 ions forming one-dimensional (1D) crystal in linear Paul traps \cite{debnath2016demonstration,friis2018observation} and over 50 ion-qubits have been used for a quantum simulation with restricted control\cite{zhang2017observation}. Extra-dimension of the crystal can provide a quadratic scaling of the number of ion-qubits in the trap.  Ion-qubits in the two-dimensional (2D) crystal intrinsically has 2D laser-induced interactions, which facilitates to study 2D many-body physics through quantum simulation such as geometric frustration, topological phase of matter, etc. \cite{bermudez2011frustrated,bermudez2012quantum,nath2015hexagonal,yoshimura2015creation,richerme2016two}. 

There have been deliberate proposals and experimental exertions to confine ions in a 2D lattice\cite{britton2012engineered,bohnet2016quantum,chiaverini2008laserless,schmied2009optimal,clark2009two,clark2011cryogenic,sterling2014fabrication,mielenz2016arrays,jain2018quantum,szymanski2012large,tanaka2014design,yan2016exploring}. In the Penning trap that uses static magnetic field and dc voltages for the confinement, hundreds of ions form a rotating 2D crystal \cite{mitchell1998direct}. Effective Ising interactions among ion-qubits in the 2D crystal have been engineered \cite{britton2012engineered}, and entanglement of spin-squeezing has been studied \cite{bohnet2016quantum}. Due to a high magnetic field condition and the fast rotation of ions in Penning traps, however, no clock-state of an ion can represent an effective spin, and it is challenging to implement individual spin controls with laser beams. 

Paul traps do not require high magnetic fields for confinement of ions and can be an alternative platform to implement 2D crystal. The main difficulty in Paul trap for producing 2D crystal of ions for quantum computation or quantum simulation lies in the existence of micromotion \cite{berkeland1998minimization} synchronous with the oscillating electric field that introduces phase modulations on laser beams for cooling and coherent operation. The micromotion can be nullified at a point or in a line, but not in a plane. To address the micromotion problem, arrays of micro traps have been proposed \cite{chiaverini2008laserless,schmied2009optimal} and small scale of arrays traps have been implemented \cite{sterling2014fabrication,mielenz2016arrays}. Due to relatively large distances between micro traps, however, the Coulomb-coupling strength between ion-qubits in different traps would be relatively weak and so as the effective spin-spin interactions induced by laser beams of coherent operations \cite{welzel2011designing,wilson2014tunable,hakelberg2019interference}. Alternatively, it has been proposed to produce a trap, where the direction of micromotion is perpendicular to the net-propagation direction of laser beams for coherent control \cite{yoshimura2015creation,richerme2016two}. Such a trap structure is relatively simple to manufacture and can easily hold tens to hundreds of ions. 

Here, we report the implementation of a Paul trap that accommodates tens of ions in a 2D crystal. The trap is a three-dimensional monolithic trap \cite{brownnutt2006monolithic,shaikh2011monolithic,wilpers2012monolithic} constructed on a single layer of gold-plated laser-machined alumina \cite{hensinger2006t,madsen2006advanced}. We carefully design the structure of the trap electrodes to precisely control the orientation of principle axis and ensure perpendicularity between the micromotion axis and the net-propagation direction of coherent operation Raman laser beams. The 2D crystal is located on the plane composed of axial and one of transverse axes, which is simply imaged to an electron-multiplying CCD (EMCCD) camera. We perform coherent operations on ion-qubits and study the spectrum of vibrational modes through globally addressed Raman laser beams. We measure the amplitude of micro-motion by comparing the strengths of carrier and micro-motion sideband transitions with three ions, where the modulation index similar to that of single ion. The strengths of effective spin-spin couplings can be similar to the values in current linear traps, which is a favorable condition for 2D quantum simulation. 

\begin{figure}[!htb]
\centering 
\includegraphics[width=0.48\textwidth]{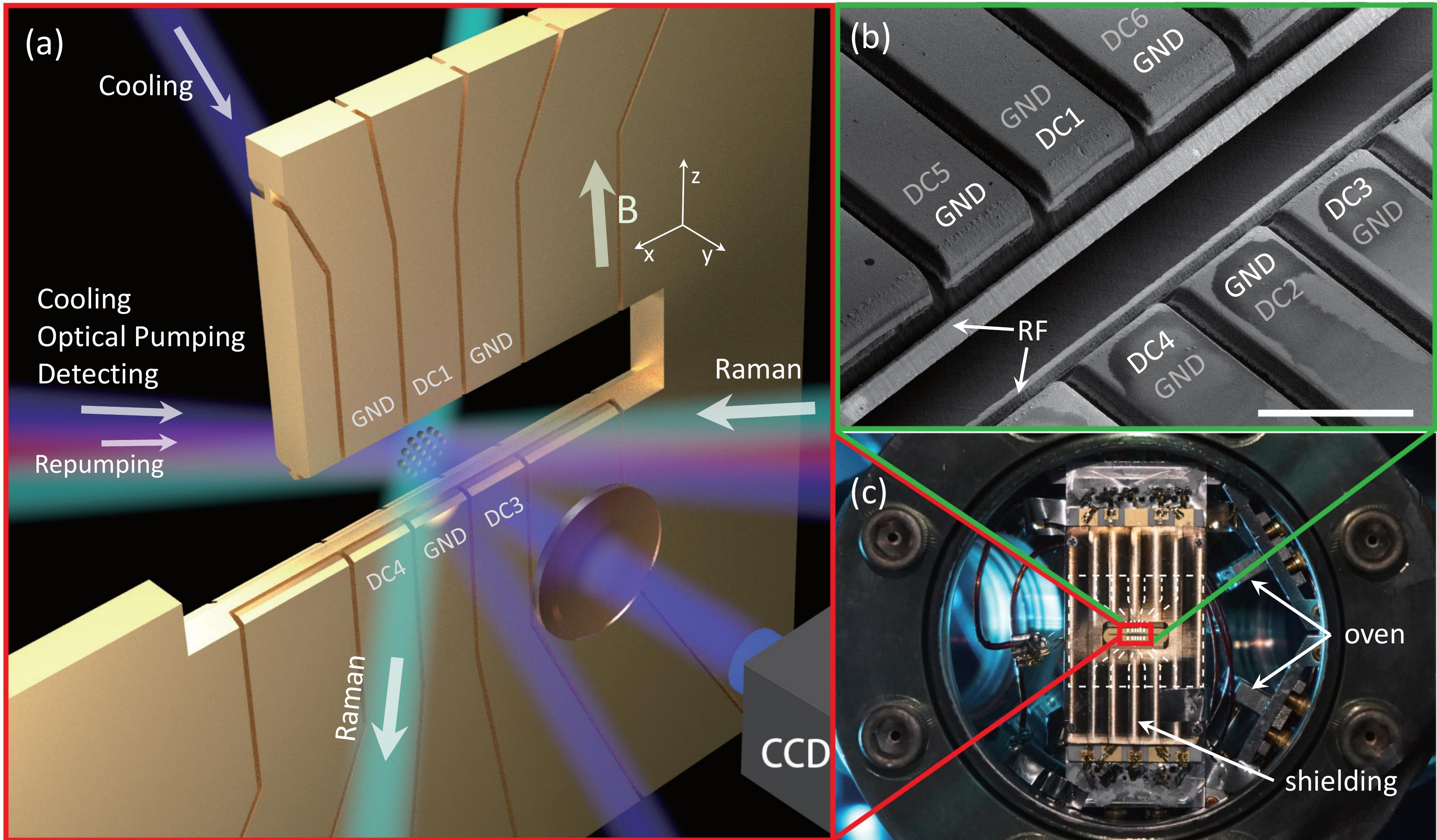}
\caption{\textbf{Trap structure and beam configuration.} (\textbf{a}) Conceptual drawing of our trap and the configuration of laser beams for cooling, pumping, detection (370 nm and 935 nm) and coherent operations (355 nm). The magnetic field is applied vertically. (\textbf{b}) Scanning electron microscope (SEM) image of our trap (white scale bar, 400 $\mu$m). The trap structure is laser-machined on a single piece of alumina with less than 10 $\mu$m precision. Gold is electro-plated on the surface of alumina with 10 $\mu$m thickness. The trap has a total of 20 electrodes, where fourteen of them are connected to GND and the others to DC sources. The gray letters label the electrodes on the opposite side of the trap.(\textbf{c}) Image of our monolithic trap mounted in a hemi-sphere vacuum chamber. The trap is shielded with stainless steel plates on the front and back, respectively, which are connected to GND. The dashed lines show the electrodes underneath of the shield. }\label{fig:trap}
\end{figure}

Our trap is fabricated in a single piece of alumina plate with gold-coating, with more details in Appendix B. Fig. \ref{fig:trap} shows the structure of our trap. The trap is monolithic and functionally separated into three layers, where front and back layers contain dc electrodes, and the middle layer is used for RF electrode as conceptually shown in Fig. \ref{fig:trap}(a). The RF electrode has a slope with the angle of 45$^{\circ}$ relative to the normal direction of the alumina piece. In each DC layer, there are ten electrodes, five electrodes on both upside and downside with a 50 $\mu$m spacing. At the center of the trap, there is a 260 $\mu$m $\times$ 4 mm slot, where ions are trapped. The Fig. \ref{fig:trap}(b) shows front side of the trap. The angle of the slope and the gap between DC and RF electrodes are optimized to maximize the trap frequency (see Appendix A for the design consideration). We use CPO (Charged Particle Optics) software to calculate the electric potential from the electrodes. We also compare the simulated potential with the real potential to calibrate the simulation coefficient for further trap simulation (see Appendix D).  In the experiment, only six of twenty electrodes are connected to the stable DC sources, and the others to GND, as shown in Fig. \ref{fig:trap}(b). 

The monolithic trap is located in a vacuum chamber shown in Fig \ref{fig:trap}(c). The trap and vacuum system is designed to ensure sufficient optical accesses. \Yb ions are loaded to the middle of the trap by photo-ionization and Doppler cooling \cite{olmschenk07manipulation}. We create the 2D crystal of ions in a plane that consists of the axial axis (x-axis) and one of the radial axes (z-axis). We apply two Doppler-cooling laser beams to couple all the three directions of ion motions, as shown in Fig. \ref{fig:trap}(a). The magnetic-field insensitive states of \Yb ion in the ground-state manifold $^2S_{1/2}$, $\ket{F=0,m_F=0}$ and $\ket{F=1,m_F=0}$ are mapped to qubit state $\ket{0}$ and $\ket{1}$, respectively. The state of the qubit is detected by the laser beam resonant to the transition between $F=1$ of $^2S_{1/2}$ and $F=0$ of $^2P_{1/2}$ and initialized to $\ket{0}$ by applying the optical pumping laser beam resonant with the transition between $F=1$ of $^2S_{1/2}$ and $F=1$ of $^2P_{1/2}$. The qubit is coherently manipulated by a pair of 355 nm picosecond pulse laser beams with beatnote frequency about the qubit transition $\omega_0=2\pi\times12.642821~{\rm GHz}$. 


We rotate the principle axes of pseudo-potential in the y-z plane by adjusting voltages $V_{\rm C}$ and $V_{\rm NC}$ on both of the center electrodes $\rm DC_{C}$ (DC1, DC2 in Fig. \ref{fig:trap}(b)) and all of the next to the center electrodes $\rm DC_{NC}$ (DC3, DC4, DC5, and DC6 in Fig. \ref{fig:trap}(b)), respectively. 
The total pseudo-potential with voltages of $V_{\rm C}$, $V_{\rm NC}$ and $V_{\rm RF}$ is described by 
\begin{equation}
\phi(x,y,z) = V_{\rm C} \phi_{\rm C}+V_{\rm NC} \phi_{\rm NC}+V_{\rm{RF}} \phi_{\rm{RF}},
\label{eq:potential}
\end{equation}
where $\phi_{\rm C}$ and $\phi_{\rm NC}$ are electric potentials at the position of $(x,y,z)$ generated by $\rm DC_{C}$ and $\rm DC_{NC}$ electrodes with unit voltage. And $\phi_{\rm RF}$ is the pseudo-potential generated by the RF electrode with root-mean-square voltage of 1 V. In y-z plane, the symmetric RF pseudo-potential can be broken by DC potentials, which leads to a elliptical total potential $\phi(x,y,z)|_{x=0}$. 
The two axes of the elliptical potential are the principle axes. In order to rotate the principle axes to y axis and z axis, we need to satisfy
\begin{eqnarray}
\partial\phi(0,y,\delta z)/\partial y|_{y=0} = 0, 
\label{eq:rotation}
\end{eqnarray}
where $\delta z$ is related to the size of 2D crystal and small enough to be in harmonic regime for our consideration. Noticing $\partial\phi_{\rm{RF}}(0,y,\delta z)/\partial y|_{y=0} = 0$ is always true, we can calculate the solution of $V_{\rm NC}/V_{\rm C}$, to satisfy Eq. (\ref{eq:rotation}) based on numerical simulation. In our trap, $V_{\rm NC}/V_{\rm C} \approx 5.11$. We should also notice that when ever we set $V_{\rm NC}/V_{\rm C}$ to the right value and rotate the principle axes to y axis and z axis, $V_{\rm RF}$ will no longer affects the rotation of the principle axes.

\begin{figure}[!htb]
\center{\includegraphics[width=0.48\textwidth]{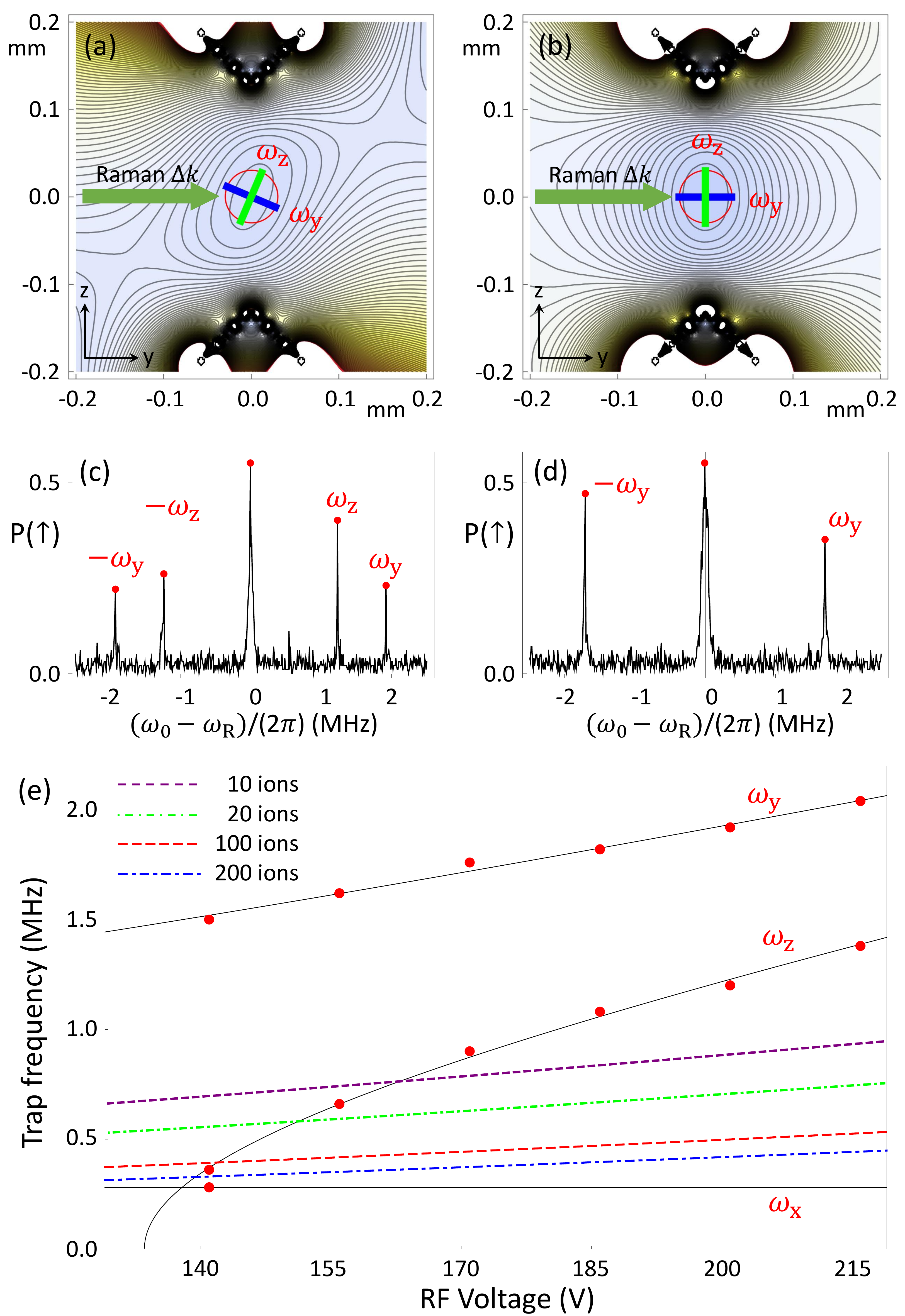}}
\caption{\label{fig:PrinAxis}\textbf{Principle axis rotation} (\textbf{a}) The contour plot of pseudo-potential when $V_{\rm NC}/V_{\rm C} =\infty$, central electrodes are connected to GND. (\textbf{b}) The contour plot of pseudo-potential when the principle axes are overlapped with y and z axes, where the voltage ratio is $V_{\rm NC}/V_{\rm C}=5.11$. (\textbf{c}) The Raman spectrum with principle axes in the condition of (a), where we can see both of the transverse modes. (\textbf{d}) The Raman spectrum with principle axes in the condition of (b). In this situation, the Raman beams can only drive the mode of the y-axis, not that of the z-axis. (\textbf{e}) Relation between two radial-mode frequencies and the RF voltages. By merely changing the RF voltage, we can realize different ratios of trap frequencies. The red dots are the experimental data, the dark lines are the fitting results. The dashed lines, which are calculated by $\omega_y/(2.264N)^{1/4}$ for different RF voltage, are the up bounds of the region where the symmetric 2D crystal can be formed for different numbers of ions.}
\end{figure}

We numerically calculate $\phi_{\rm C}$, $\phi_{\rm NC}$ and $\phi_{\rm RF}$ with CPO software. We set the RF signal to be $\omega = 2 \pi \times 40$ MHz and $V_{\rm{RF}} = 80$ V. When $V_{\rm NC}/V_{\rm C}=\infty$ with $V_{\rm NC}= 1.5$ V, vertical principle axis (green line in Fig. \ref{fig:PrinAxis}(a)) is clockwise rotated by 22.9$^{\circ}$ from the z-axis. When the ratio $V_{\rm NC}/V_{\rm C}=0$ with $V_{\rm C}= 1.5$ V, the green axis is counter-clockwise rotated by 5.7$^{\circ}$ from the z-axis. As shown in Fig. \ref{fig:PrinAxis}(b), when the ratio $V_{\rm NC}/V_{\rm C}=5.11$, the green axis is in line with z-axis. 

We experimentally confirm the rotation of the principle axes in y-z plane with single ion by observing the disappearance of the Raman coupling to z-axis vibrational mode. The spectrum of vibrational modes, as shown in Fig. \ref{fig:PrinAxis}(c)(d) is measured by the following procedure: 1) we perform Doppler cooling on ion-crystal, which results in thermal states with $\bar{n}\approx 7.1$, Doppler cooling limit, and initialize the internal states to $\ket{\downarrow}$ by applying the standard optical pumping technique. 2) We apply Raman beams with a net $k$-vector perpendicular to the z-axis. Once the beatnote-frequency $\omega_{\rm R}$ of Raman beams matching to $\omega_{\rm 0} \pm \omega_{\rm y, z}$, sideband transitions occurs \cite{leibfried2003quantum}, which can be detected by the fluorescence of ions that is collected by imaging system and PMT (Photo-multiplier tube). In Fig. \ref{fig:PrinAxis}(c), the voltage ratio is close to the condition of $V_{\rm NC}/V_{\rm C}=\infty$ in Fig. \ref{fig:PrinAxis}(a), where the principle axes are tilted away from y-z axes. The net $k$-vector of Raman beams is along the y-axis, which can excite both directions of vibrational modes. Thus, two peaks in blue-sidebands $\left(\delta= \omega_{\rm y, z}\right)$ as well as red-sidebands $\left(\delta= -\omega_{\rm y, z}\right)$ are clearly visible in Fig. \ref{fig:PrinAxis}(c), where detuning $\delta=\omega_{\rm R} -\omega_{\rm 0}$. However, when the principle axes are rotated to y-z axes as shown in Fig. \ref{fig:PrinAxis}(b), Raman beams cannot excite the vibrational mode along z-axis, which results in vanishing a peak in the Raman spectrum. Based on the spectrum of Fig. \ref{fig:PrinAxis}(d), we estimate that deviation of the principle axes from y-z axes is below $0.40^{\circ}$.

In order to produce a 2D ion crystal in z-x plane, we need to satisfy $\omega_{\rm y} > (2.264 N)^{1/4} \omega_{\rm x,z}$ \cite{richerme2016two,dubin1993theory}. First, keeping the principle axes to y-z axes, we can calculate the voltage solution for DC electrodes with a given axial trap frequency $\omega_{\rm x}$. With determined DC potential, the relation between $\omega_{\rm y}$ and $\omega_{\rm z}$ is given by \cite{leibfried2003quantum}
\begin{eqnarray}
\omega_{\rm y}^2-\omega_{\rm z}^2 = C V_{\rm NC},
\label{eq:difference}
\end{eqnarray}
(see Appendix C) where $C$ is a positive constant determined by the trap geometry. In the case of $V_{\rm{RF}} = 0$, the z-axis potential, the shallower potential respective to that of the y-axis according to Eq. (\ref{eq:difference}), becomes anti-harmonic, which indicates $\omega_{\rm z}^2 <0$ and $\omega_{\rm y}^2 < C V_{\rm NC}$. On the other hand, since $\omega_{\rm y} and $ $\omega_{\rm z}$ are monotonously increase with $V_{\rm{RF}}$, there is a critical value of $V_{\rm RF}$ that makes $\omega_{\rm y}^2 = C V_{\rm NC}$ and $\omega_{\rm z}^2 = 0$. Therefore, we can tune $\omega_{\rm y}$ from $\sqrt{CV_{\rm NC}}$ to $\infty$, $\omega_{\rm z}$ from near zero to $\infty$ by tuning $V_{\rm RF}$. As shown in Fig. \ref{fig:PrinAxis}(e), with different values of $V_{\rm{RF}}$, we can have $\omega_{\rm z}/\omega_{x}$ from 0 to 2.72 for 10 ions to realize 2D ion crystal with different aspect ratios.

\begin{figure}[!htb]
\center{\includegraphics[width=0.48\textwidth]{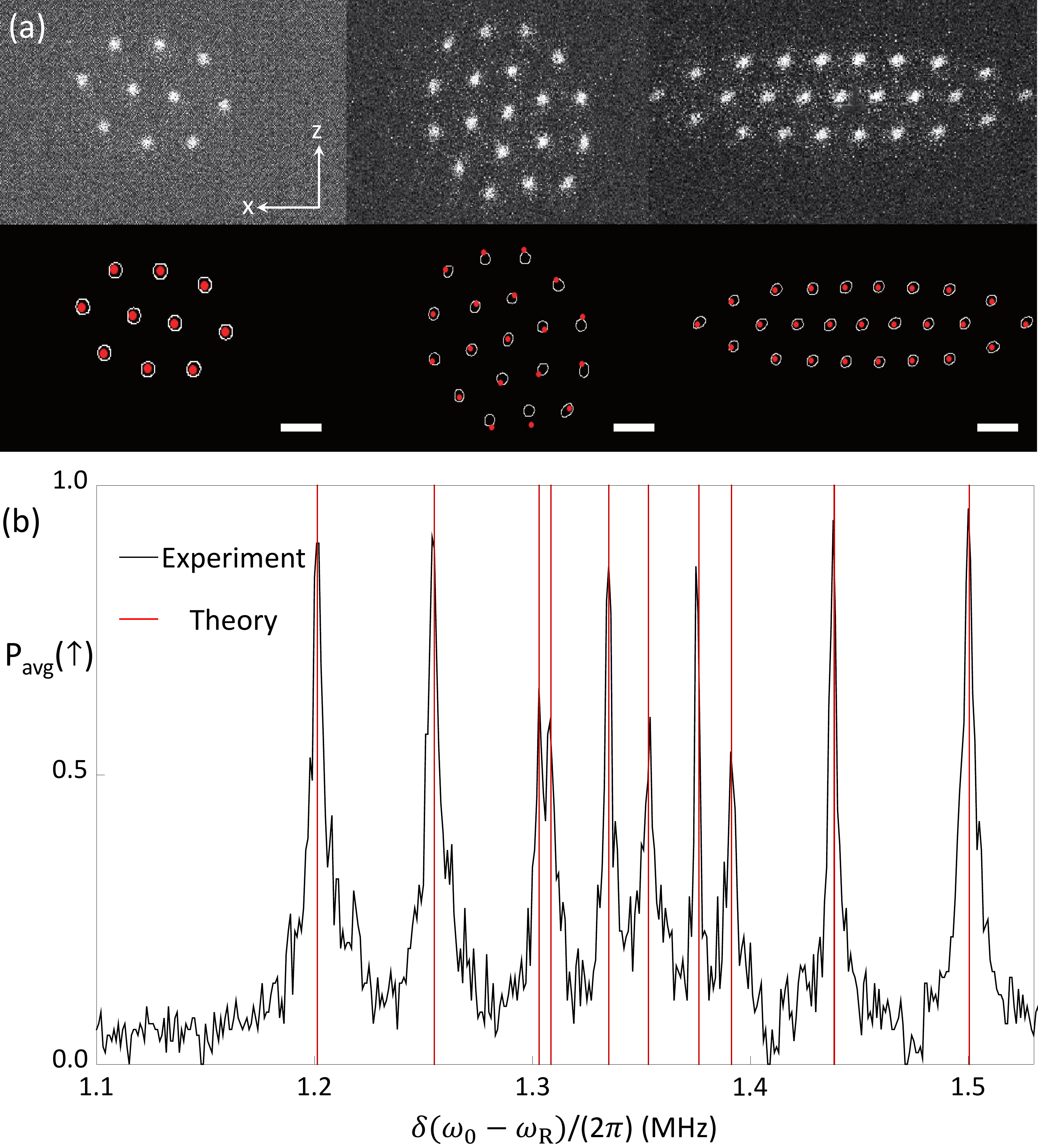}}
\caption{\label{fig:2D_crystal}\textbf{Geometry and mode structure of 2D-ion crystals.} (\textbf{a}) CCD images of 10,19, and 25 ion crystals with different trap frequencies. The above ones are raw data taken from EMCCD, and the bottom ones compare the numerical simulation and the real data where the white cycles are the positions of ions read out from the CCD pictures and the red points are simulation results. The white scale bars represent 5 $\mu$m distance. (\textbf{b}) Raman spectrum of the 2D crystal with ten ions. The black curve is the experiment result and the red lines are the theoretical prediction of the mode frequencies \cite{richerme2016two}.}
\end{figure}

Once the requirements of principle axes and trap-frequencies for 2D crystal are satisfied as discussed above, we can confine ions in the xz-plane. Fortunately, the strongest trap frequency in our monolithic trap is in y-axis due to the geometry of the trap, which allows us to easily image the 2D crystal with the same imaging system to 1D chain. The fluorescence of ions in 2D crystal can be directly imaged through an objective lens to CCD camera as shown in Fig. \ref{fig:trap}(a). Figure \ref{fig:2D_crystal}(a) are the images of the 2D crystals and demonstrate the control capability for shapes of 2D crystals with various settings of trap frequencies. 
For the image of 10 ions, the trap frequencies are  $\{\omega_{\rm x},\omega_{\rm y},\omega_{\rm z}\}/(2\pi)=\{0.427, 1.5, 0.561\}$MHz. 
For the image of 19 ions and 25 ions the trap frequencies are $\{\omega_{\rm x},\omega_{\rm y},\omega_{\rm z}\}/(2\pi)=\{0.28, 1.50, 0.26\}$MHz and $\{\omega_{\rm x},\omega_{\rm y},\omega_{\rm z}\}/(2\pi)=\{0.28, 1.63, 0.68\}$MHz respectively. For 25 ions, these frequencies look violating the bound for the 2D crystal in Fig. \ref{fig:PrinAxis}(e). However, it is a complicated situation since one of the frequencies, $\omega_{\rm x}$, is still below the bound. We numerically study the situation carefully and discuss in the Appendix E. The geometries of the crystal are in agreement with the numerical simulation.  We simulate the geometry configuration of the ion crystal by numerically minimizing the electrical potential of the ions in a three dimensional harmonic trap \cite{richerme2016two}.





We first verified the dimension of the crystal by imaging the crystal, but it is hard to distinguish whether there are any ions out of the ion crystal plane by an objective lens with finite depth of field. The dimension of the crystal is further verified by measuring the transverse mode structure. Similar to the single ion case, we drive the different transverse modes of a 10-ion crystal by varying the detuning between Raman beams. Fig \ref{fig:2D_crystal}(b) shows the resulting spectrum, where each peak represent a motional mode in the y-axis. The mode spectrum is consistent with the theoretical simulation based on trap frequencies and geometry of 2D ion crystal\cite{richerme2016two}. Similar to linear chain case\cite{enzer2000observation}, when the phase transition from a 2D crystal to a 3D crystal happens the minimal frequency of the mode along the y-axis will tend to be negative. Our measured mode frequencies are far away from zero, which confirm the dimension of the crystal is two.


\begin{figure}[!htb]
\center{\includegraphics[width=0.48\textwidth]{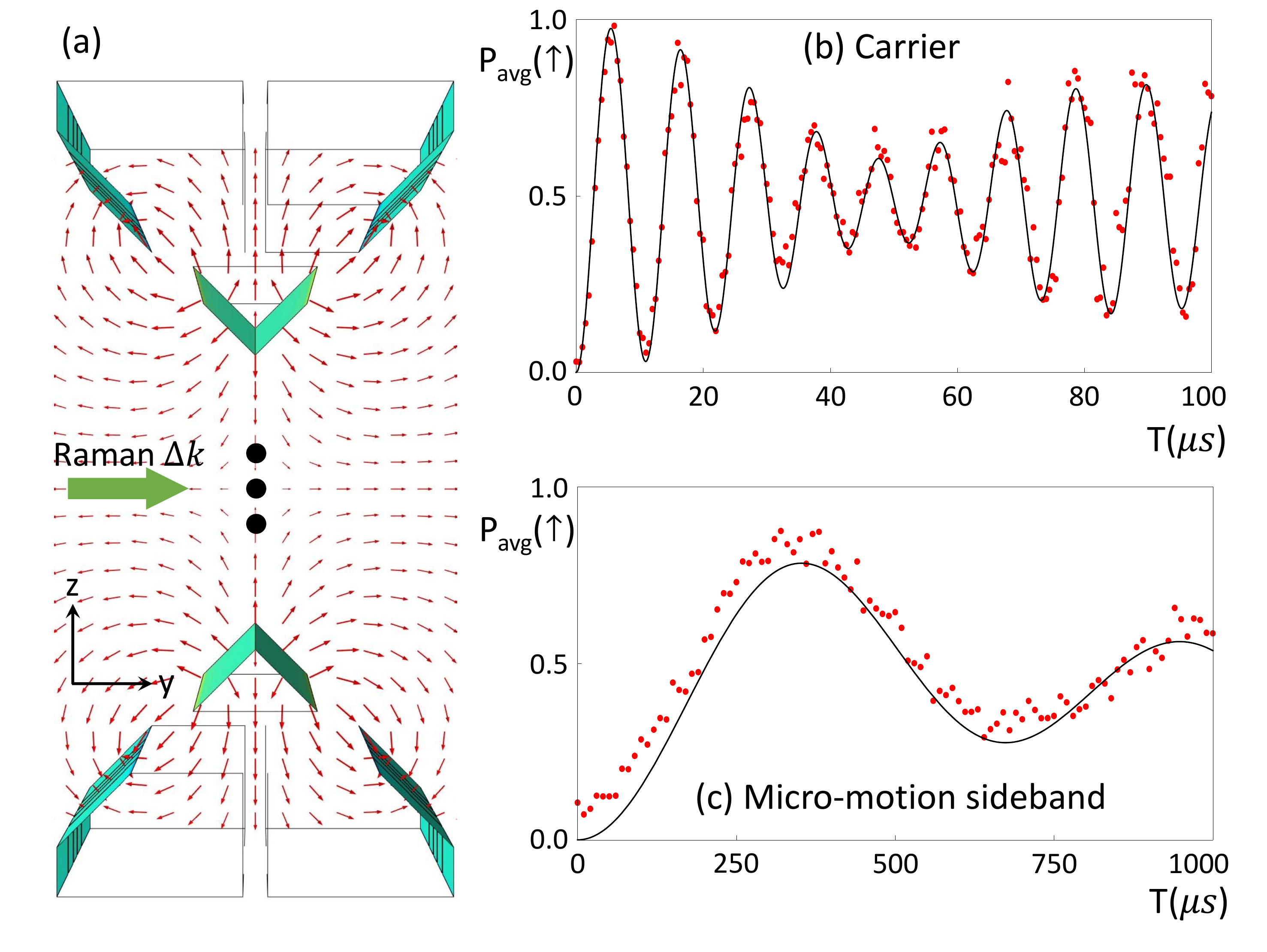}}
\caption{\label{fig:Micromotion}\textbf{Micro-motion direction and strength in the trap.}(\textbf{a}) Vector plot of the RF field is done by CPO. The simulation shows if the crystal is located in the z-x plane, the direction of micro-motion is along the z-axis and perpendicular to the y-axis which is the net propagation direction of Raman beams. (\textbf{b}) Carrier transition of three ions in triangular crystal detected by PMT. The beating signal comes from the unbalanced Rabi-frequency of each ion. (\textbf{c}) Micro-motion transition of the three ion crystal after compensating the micro-motion.}
\end{figure}

Ideally, if the crystal is located in the z-x plane, the direction of micro-motion is along the z-axis and perpendicular to the y-axis, which is the net propagation direction of Raman beams. In practice, there are two possible imperfection sources that make the crystal deviate from the ideal micro-motion condition: 1) Stray electric field, which induces displacement; 2) Fabrication imperfection of the electrodes, which induces the tilt around the z-axis.  
To minimize the micro-motion from theses sources, we first compensate the straight field with a single-ion, then mitigate the tilt errors by slightly rotating the crystal. With the single ion, the micro-motion compensation is done by overlapping the position of the ion to the null point of the RF electric field \cite{berkeland1998minimization}. We first compensate the extra-field in z direction by changing the voltage of $\rm \{DC2,DC3,DC4 \}$ or $\{$DC1,DC5,DC6$\}$ simultaneously with the ratio  $\{$1,5.11,5.11$\}$, which is able to keep the principle axes direction and avoid generating the displacement along y axis. We can also change the voltage of electrodes $\{$DC1,DC3,DC4$\}$ or $\{$DC2,DC5,DC6$\}$ with ratio $\{$1,5.11,5.11$\}$ to compensate the extra-field in y direction. For the z-axis compensation,  we minimize the change of ion position depending on RF power and for the y-axis compensation, we minimize the micro-motion sideband transition of Raman beams.  
For the error induced by the fabrication imperfection, we slightly change the voltage of electrodes $\{$DC3,DC4,DC5,DC6$\}$ with ratio $\{$1,1,1,1$\}$ to rotate the crystal around x axis and with ratio $\{$1,-1,1,-1$\}$ to rotate the crystal around z axis. With the control, we also minimize the Rabi-frequency of the micro-motion sideband transition with three ions.


The strength of the micro-motion is quantified by measuring the ratio between two Rabi frequencies of the carrier and the micro-motion transition\cite{berkeland1998minimization}. We measure the micro-motion strength in a three-ion 2D crystal. We first sequentially  apply Doopler cooling and EIT cooling \cite{eitcooling} to cool the 2D crystal down to near the motional ground state. Then we drive the Rabi flopping and measure the Rabi frequency of the carrier and the micro-motion sideband transition. For each flopping, we collect the overall counts of three ions with PMT and fit the result with three Rabi frequency. The fitting gives us three carrier $\pi$-time $\{$5.96, 5.40, 5.19$\}\mu$s and three micro-motion sideband $\pi$-time $\{$474, 440, 317$\}\mu$s. The modulation index, which is given by $\beta/2 = \Omega_{\text{micro}}/\Omega_{\text{carrier}}$, has a maximun possible value of $0.038$ and a minimal possible value of $0.021$, which are similar to single ion situation.

We also experimentally study the heating of the vibrational modes in our trap with a single ion. We first prepare the ground-state of radial vibrational modes by Raman-sideband cooling, wait for a certain duration and measure average phonon-number $\bar{n}$ for the mode of interest. We estimate $\bar{n}$ by Fourier transforming the blue-sideband transitions \cite{leibfried2003quantum}. We find that the heating rate of y-axis mode with the principle axes of 2D crystal (Fig. \ref{fig:PrinAxis}(b)) is around 360 quanta per second, which is about 2.5 times larger than that with the condition of Fig. \ref{fig:PrinAxis}(a). It is understandable, since the noise of environmental electric field along y-axis would be more severe than those of the other axes. 

Our trap can be considered as an ideal platform for implementing various proposals of quantum simulations with 2D crystal \cite{bermudez2011frustrated,bermudez2012quantum,nath2015hexagonal,yoshimura2015creation,richerme2016two}. It can be used to observe a structural quantum phase transition from 2D to 3D with relaxed requirements\cite{morigiyoutube}. Incorporating the capability of individual control and detection, universal quantum computation also can be achieved with more number of qubits than in a linear chain. The capability of laser addressing in a two dimensional space has been demonstrated with different techniques\cite{crain2014individual,mcgloin2003applications}. 
The detection of individual ions already has been well established using camera with high detection efficiency \cite{myerson2008high}. 

Furthermore, the 2D crystal is a natural platform for the fault-tolerant quantum computation schemes with 2D geometry, including the surface code\cite{bombin2006topological}, the Bacon-Shor code\cite{aliferis2007subsystem} and the (2+1) dimensional fault-tolerant measurement-based quantum computing\cite{raussendorf2007topological,bombin20182d,newman2019generating}. The full connectivity of trapped-ion system provides the capability of implementing any fault-tolerant scheme without extra overhead at the circuit level\cite{li20192d}. However, to implement a 2D topological code on an 1D ion chain, one has to map some local interactions to long distance gates or shuttle the ions, which requires longer gate time\cite{bermudez2017assessing,trout2018simulating,blumel2019power}. With a 2D ion crystal, the locality of 2D topological codes can be preserved, without the loss of full connectivity.

\section*{Acknowledgments}

We thank Jincai Wu and Haifeng Zhu at Interstellar Quantum Technology (Nanjing) Ltd for many useful discussions on fabrication technique. This work was supported by the National Key Research and Development Program of China under Grants No. 2016YFA0301900 and No. 2016YFA0301901 and the National Natural Science Foundation of China Grants No. 11374178, No. 11574002, and No. 11974200.

Y.W. and M.Q contributed equally to this work.

\section{APPENDIX}
\section{\label{sec:simulation} APPENDIX A: STRUCTURE OF THE TRAP}
\begin{figure}[!htb]
\center{\includegraphics[width=0.48\textwidth]{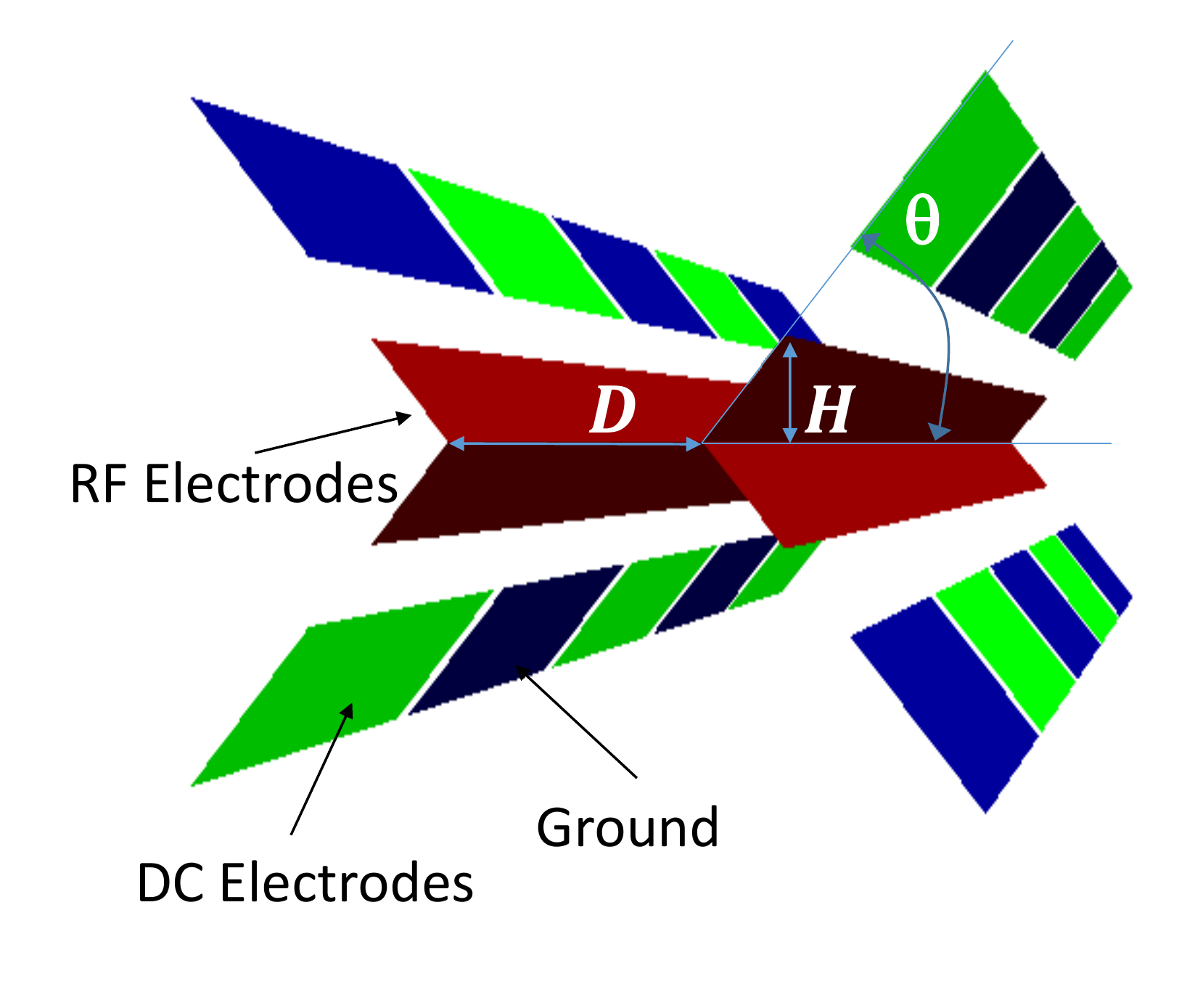}}
\caption{\label{fig:CPO}\textbf{Important geometric parameters for the trap design }. We generate 3D models with all combinations of three parameters and calculate the pseudo-potential and the secular frequency with CPO software. Maximizing the secular frequencies given fabrication limitation and laser-light scattering, we choose $D = 260$ $\mu$m, $\theta = 45^\circ$ and $H = 40$ $\mu$m for the trap.}
\end{figure}

We use CPO software to simulate the trap performance with various geometric parameters. There are three important parameters for the trap design: the distance between two RF electrodes $D$, the height of RF electrodes $H$, and the angle of the slope $\theta$ as shown in Fig \ref{fig:CPO}. We optimize these three parameters mainly to achieve large secular frequencies in the radial direction given fabrication limitation. 
The secular frequency is approximately inverse-proportional to $D^2$\cite{leibfried2003quantum}, which is inspected in our numerical simulation. We balance the requirement of large trap frequency and low UV-light scattering, which leads to the choice of $D = 260$ $\mu$m. 
For the slope angle $\theta$, our simulation shows the best performance at $\theta \approx 47^\circ$. Due to the fabrication difficulty of the angle, we choose $\theta = 45^\circ$. Our simulation shows the best value of $H$ is around $30$ $\mu$m. Considering the laser cutting precision, we decide $H = 40$ $\mu$m.

\section{\label{sec:fabrication}APPENDIX B: FABRICATION PROCESS}
\begin{figure}[!htb]
\center{\includegraphics[width=0.48\textwidth]{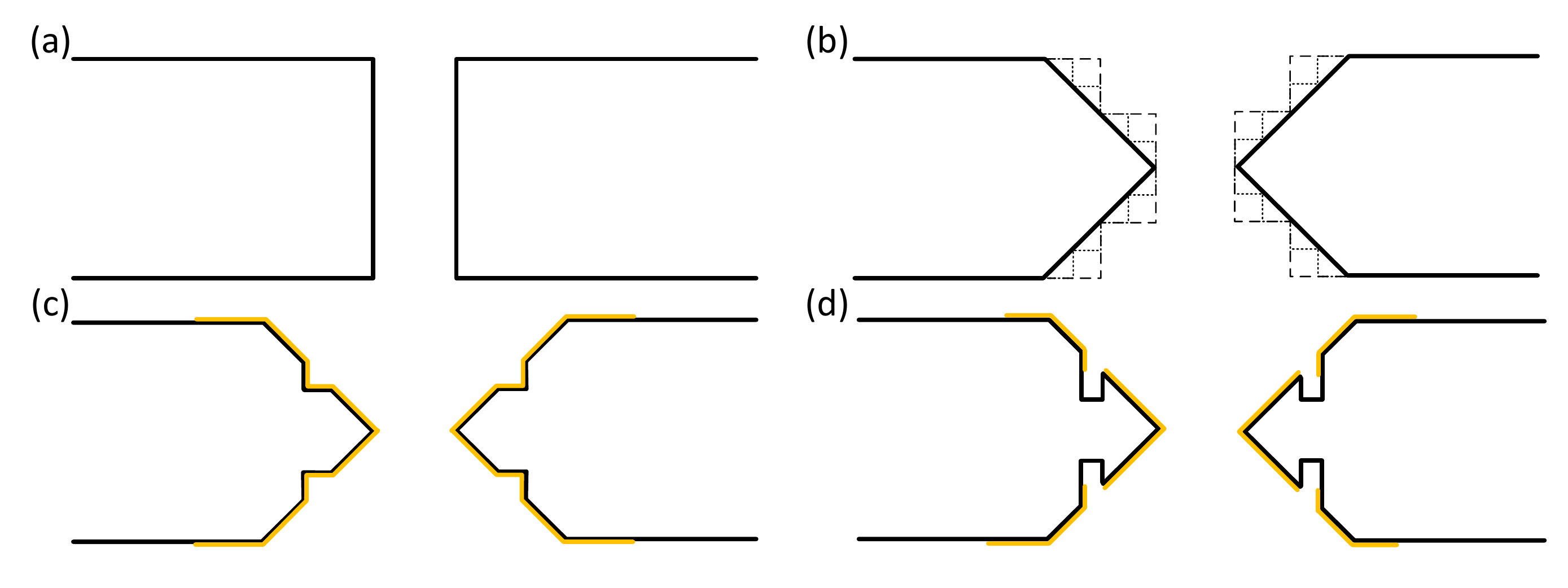}}
\caption{\label{fig:fabrication}\textbf{Steps for fabricating the structure of electrodes. }(\textbf{a}) Laser cut the 260 $\mu$m slot. (\textbf{b}) Cut 40 small steps for each slope with 45$^{\circ}$. (\textbf{c}) Laser cut the small groove and electroplate gold on the surface. (\textbf{d}) Cut the slots on the grooves and two sides of the chip to electrically separate all DC and RF electrodes. }
\end{figure}

The substrate is a single piece of alumina with the thickness of 380 $\mu$m and the surface flatness of less than 30 nm. The electronic structure is fabricated by the laser-machining and coated with 3 $\mu$m gold by electroplating technology. The detailed procedure to fabricate the electrodes structure is as follows: 1) Carve a slot of 260 $\mu$m at the center of the piece, as shown in Fig. \ref{fig:fabrication}(a); 2) Make a slope of 45$^{\circ}$ on each side by cutting small steps to fit the slope as shown in Fig. \ref{fig:fabrication}(b); 3) Make a tiny groove on each slope. The width of the groove is around 50 $\mu$m; 4) Do gold coating on both sides of the chip as shown in Fig. \ref{fig:fabrication}(c); 5) Cut deeper in the groove position to remove gold. The center layer is electrically separated with top and bottom layer; 6) Laser cut the slots on top and bottom layer to electrically separate all DC electrodes. Among all the steps, the second is the subtlest one. The geometry of the four slopes is crucial for the ion control with DC voltages. In step 2), for each slope, we apply 40 times of laser cutting with different duration and 5 $\mu$m shift on cutting position. The cutting duration for each pulse is calculated base on the calibrated relationship between the cutting depth and the cutting time. The laser cutting precision is $\pm$1 $\mu$m, which is limited by the worktable instability. Using a laser with the power of 2W, the wavelength of 355 nm and the beam waist of around 15 $\mu$m, we can have the cutting speed to be 100 mm/s.

\section{\label{sec:calculation}APPENDIX C: TRAP FREQUENCY CALCULATION}

According to Ref. \cite{leibfried2003quantum}, we can write the time dependent potential of the trap as follows:
\begin{eqnarray}
\label{eq:potentialCal}
    \phi(x,y,z,t) &= \sum_{\rm E \in \rm DC}\frac{1}{2}V_{\rm E}(\alpha_{\rm E} x^2+\beta_{\rm E} y^2 +\gamma_{\rm E} z^2)\\\nonumber
    &+V_{\rm RF} \cos(\omega_{\rm RF}t)(\alpha'x^2+\beta'y^2+\gamma'z^2),
\end{eqnarray}
where $V_{\rm E}$ is the voltage applied on the DC$_{\rm E}$ electrode, $\alpha_{\rm E}$, $\beta_{\rm E}$, $\gamma_{\rm E}$ are geometric factors determined by the geometry of the DC$_{\rm E}$ electrode, $V_{\rm RF}$ is the root mean square of the voltage applied on RF electrode, $\alpha'$, $\beta'$ and $\gamma'$ are geometric factors determined by the geometry of the RF electrode. We note that x,y and z axes in the Eq. (\ref{eq:potentialCal}) should be three principle axes of the trap potential. All the geometric factors will change based on the different rotation of the principle axes. The condition that the potential has to fulfill the Laplace equation $\Delta \Phi = 0$ leads to the restrictions as follows:
\begin{align}
    &\alpha+\beta+\gamma = 0, \\
    &\alpha'+\beta'+\gamma' = 0.
\end{align}
With our symmetric RF electrodes in the axial direction, it's clear that $\alpha' = 0$, which leads to $\beta' = -\gamma'$. Solving the Mathieu equation for three directions, we can have the results as:
\begin{align}
    \omega_{\rm x} = \sqrt{\frac{4e\sum_{\rm E \in \rm DC}V_{\rm E}\alpha_{\rm E}}{m \omega_{\rm RF}^2}+\frac{2e^2V_{\rm RF}^2\alpha'^2}{m^2\omega_{\rm RF}^4}}\frac{\omega_{\rm RF}}{2},\\
    \omega_{\rm y} = \sqrt{\frac{4e\sum_{\rm E \in \rm DC}V_{\rm E}\beta_{\rm E}}{m \omega_{\rm RF}^2}+\frac{2e^2V_{\rm RF}^2\beta'^2}{m^2\omega_{\rm RF}^4}}\frac{\omega_{\rm RF}}{2},\\
    \omega_{\rm z} = \sqrt{\frac{4e\sum_{\rm E \in \rm DC}V_{\rm E}\gamma_{\rm E}}{m \omega_{\rm RF}^2}+\frac{2e^2V_{\rm RF}^2\gamma'^2}{m^2\omega_{\rm RF}^4}}\frac{\omega_{\rm RF}}{2}.
\end{align}
Due to $\beta' = - \gamma'$, we can have
\begin{align}
    \omega_{\rm y}^2-\omega_{\rm z}^2 = \frac{e}{m}[\sum_{\rm E \in \rm DC}(\beta_{\rm E}-\gamma_{\rm E}) V_{\rm E}].
\end{align}
This equation explains the Eq. (\ref{eq:difference}) in the main text. As we mentioned before, all the geometric factors are determined by the rotation of the principle axes. 

\section{APPENDIX D: TRAP SIMULATION CALIBRATION}

\begin{figure}[!htb]
\center{\includegraphics[width=0.48\textwidth]{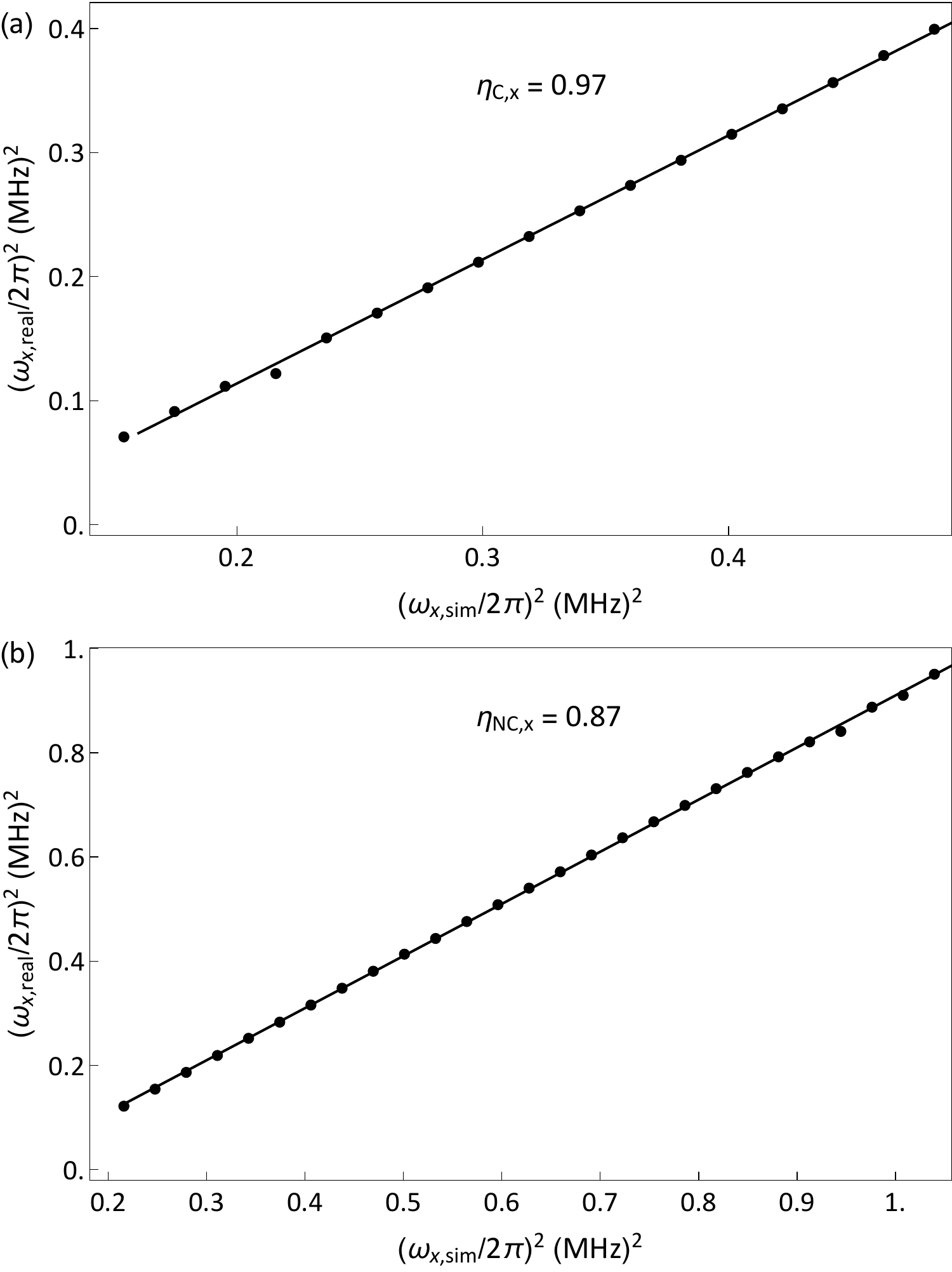}}
\caption{\label{fig:axialCali}\textbf{Axial potential calibration }(\textbf{a}) We only change the value of $V_{\rm NC}$ and measure $\omega_{\rm real, x}$. Then we simulate the ideal $\omega_{\rm sim, x}$ using the same DC voltage condition. By linear fitting the points \{$\omega_{\rm x,real}^2, \omega_{\rm x,sim}^2$\}, we can get $\eta_{\rm NC,x} = 0.87$. (\textbf{b}) We only change the value of $V_{\rm C}$ and plot all the points  \{$\omega_{\rm x,real}^2, \omega_{\rm x,sim}^2$\}. By linear fitting the points, we can get $\eta_{\rm C,x} = 0.97$.}
\end{figure}

Due to the fabrication imperfection, the real trap potential may deviate from the ideal model in simulation. We develop a method to quantitatively calibrate difference between the reality and the simulation, which is useful for the further simulation and the prediction of the trap behavior. Take $\phi_{\rm C}(x,0,0)$ as an example, we can describe difference between the reality and the simulation as follows: 

\begin{equation}
\phi_{\rm real, C}(x,0,0) = \eta_{\rm C, x} \phi_{\rm sim, C}(x,0,0),
\end{equation}
where $\phi_{\rm real, C}(x,0,0)$ is the real potential generated by electrode DC$_{\rm C}$ along the x-axis, $\phi_{\rm sim, C}(x,0,0)$ is the simulated potential, and $\eta_{\rm C, x}$ is the imperfection coefficient for DC$_{\rm C}$ in x-axis. We study the relationship between the real axial trap frequency and the simulated axial trap frequency to calibrate $\eta_{\rm C, x}$.

We start from calculating the axial mode frequency, which is $\omega_{\rm x} = \sqrt{\partial^2 \phi(x,0,0)/\partial^2x}|_{x=0}$. By using the expression of $\phi$ in Eq. (\ref{eq:potential}), we can have 

\begin{equation}
\label{eq:axialFre}
\omega_x^2 = V_{\rm C}\frac{\partial^2 \phi_{\rm C}(x,0,0)}{\partial^2x}+V_{\rm NC}\frac{\partial^2 \phi_{\rm NC}(x,0,0)}{\partial^2x}|_{x=0}.
\end{equation}
With the Eq. (\ref{eq:axialFre}) and fixed value of $V_{\rm NC}$, we can treat $\omega_x^2$ as a linear function with $V_{\rm C}$, which has the slope as $a = \frac{\partial^2 \phi_{\rm C}(x,0,0)}{\partial^2x}|_{x=0}$ and the intercept as $b = V_{\rm NC}\frac{\partial^2 \phi_{\rm NC}(x,0,0)}{\partial^2x}|_{x=0}$. We can write two version of Eq. (\ref{eq:axialFre})
\begin{align}
\omega_{\rm x,real}^2 & = a_{\rm real} V_{\rm C} +b_{\rm real} \label{eq:axialFreqReal},\\
\omega_{\rm x,sim}^2 & =  a_{\rm sim} V_{\rm C} +b_{\rm sim} \label{eq:axialFreqSim}, 
\end{align}
where
\begin{align}
a_{\rm real} = \frac{\partial^2 \phi_{\rm real,C}(x,0,0)}{\partial^2x}, \\
a_{\rm sim} =  \frac{\partial^2 \phi_{\rm sim,C}(x,0,0)}{\partial^2x}.
\end{align}
So we know
\begin{equation}
\eta_{\rm C,x} = \frac{\phi_{\rm real, C}(x,0,0)}{\phi_{\rm sim, C}(x,0,0)} = \frac{a_{\rm real}}{a_{\rm sim}}. \label{eq:axialFreqCompare}
\end{equation}
Combining Eq. (\ref{eq:axialFreqReal}), Eq. (\ref{eq:axialFreqSim}) and Eq. (\ref{eq:axialFreqCompare}), with the same value of $V_{\rm C}$, we can have
\begin{equation}
\label{eq:etaaxial}
\omega_{\rm x,real}^2 = \eta_{\rm C,x} \omega_{\rm x,real}^2+b_{\rm C},
\end{equation}
where $b_{\rm C}$ is an intercept determined by $V_{\rm NC}$ and geometries of other electrodes. We measured axial trap frequency $\omega_{\rm x,real}$ by adding a modulation signal on one of the DC electrodes and checking the ion image. When the modulation frequency is close to the axial mode frequency, the motion of the ion is resonantly excited and melting in the axial direction. By changing $V_{\rm C}$ and plotting the points \{$\omega_{\rm x,real}^2, \omega_{\rm x,sim}^2$\} in Fig. \ref{fig:axialCali}(a), we can fit the coefficient of $\eta_{\rm C,x} = 0.97$. By doing same measurement but only changing $V_{\rm NC}$, we can obtain $\eta_{\rm NC,x} = 0.87$. $\eta_{\rm C,x}$ is close to 1, which means the geometry of the center electrodes is near perfect in the axial direction. On the other side, $\eta_{\rm C,x} = 0.87$ indicate that DC$_{\rm NC}$ electrodes are further away from the ion in the reality than in the simulation. Whenever we want to simulate the axial potential, we need to include $\eta_{\rm C,x}$ and $\eta_{\rm NC,x}$ in consideration.

\begin{figure}[!htb]
\center{\includegraphics[width=0.9\textwidth]{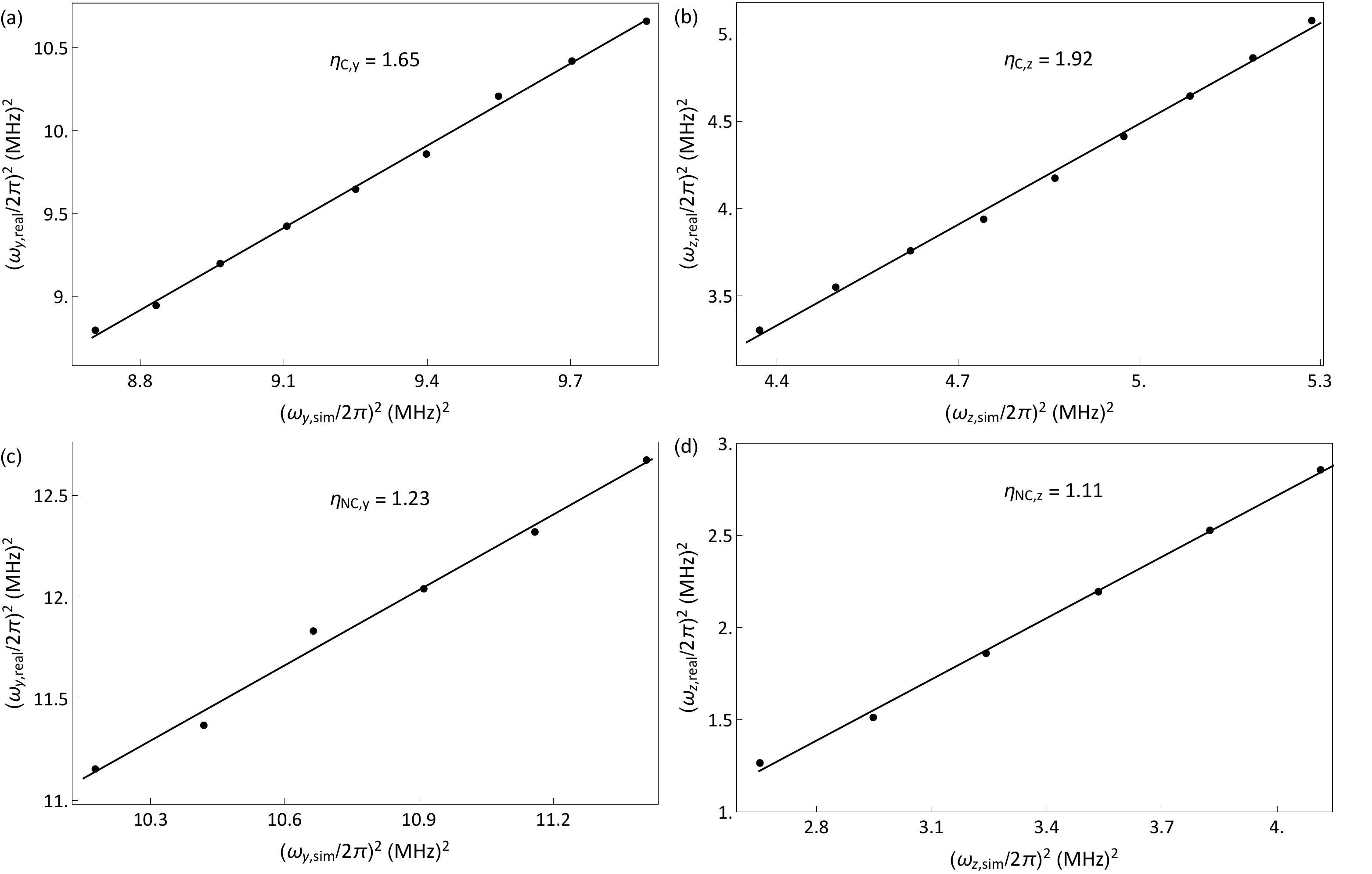}}
\caption{\label{fig:radiallCali}\textbf{Radial potential calibration }(\textbf{a}) We only change the value of $V_{\rm C}$ and measure $\omega_{\rm real, y}$. Then we simulate the ideal $\omega_{\rm sim, y}$ using the same DC voltage condition. We check the rotation of the principle axes in the experiment and the in the simulation to ensure the approximation assumption. By linear fitting the points \{$\omega_{\rm x,real}^2, \omega_{\rm x,sim}^2$\}, we can get $\eta_{\rm C,y} = 1.65$. With the same procedure in (b), (c) and (d), we calibrated the imperfection factor $\eta_{\rm C,z} = 1.92$, $\eta_{\rm NC,y} = 1.23$ and $\eta_{\rm NC,z} = 1.11$}
\end{figure}

To calibrate the imperfection coefficients of two radial principle axes, y-axis and z-axis, we execute the same procedure as the axial calibration with more careful consideration about the principle-axes rotation. During the process of changing $V_{\rm NC}$ or $V_{\rm C}$, only if we keep the rotation angle of the principle axes in a small regime, we can have the similar equations as Eq. (\ref{eq:etaaxial}) for y-axis and z-axis:
\begin{align}
\label{eq:etaradial}
\omega_{\rm y,real}^2 &\approx \eta_{\rm C,y} \omega_{\rm y,real}^2+b_{\rm C,y},\\
\omega_{\rm z,real}^2 &\approx \eta_{\rm C,z} \omega_{\rm z,real}^2+b_{\rm C,z},\\
\omega_{\rm y,real}^2 &\approx \eta_{\rm NC,y} \omega_{\rm y,real}^2+b_{\rm NC,y},\\
\omega_{\rm z,real}^2 &\approx \eta_{\rm NC,z} \omega_{\rm z,real}^2+b_{\rm NC,z}.
\end{align}
All the data is shown in Fig. \ref{fig:radiallCali}. From the data and the linear fitting, we can obtain $\eta_{\rm C,y} = 1.65$, $\eta_{\rm C,z} = 1.92$, $\eta_{\rm NC,y} = 1.23$ and $\eta_{\rm NC,z} = 1.11$. All these imperfection coefficients are larger than 1, which indicates that all,  relative to the ideal model, the DC electrodes are closer to the ion in the radial direction in the reality. When we simulate the radial potential and check the principle axes rotation in yz-plane, we use the average value $\eta_{\rm C,yz} = 1/2(\eta_{\rm C,y}+\eta_{\rm C,z}) = 1.785$ and $\eta_{\rm NC,yz} = 1/2(\eta_{\rm NC,y}+\eta_{\rm NC,z}) = 1.17$ to be the coefficients multiplied to $\phi_{\rm C}(0,y,z)$ and $\phi_{\rm NC}(0,y,z)$.

\onecolumngrid

\twocolumngrid

\section{APPENDIX E: GEOMETRY OF ION CRYSTAL AND SIMULATION OF MODE FREQUENCIES}

\begin{figure}[!htb]
\center{\includegraphics[width=0.9\textwidth]{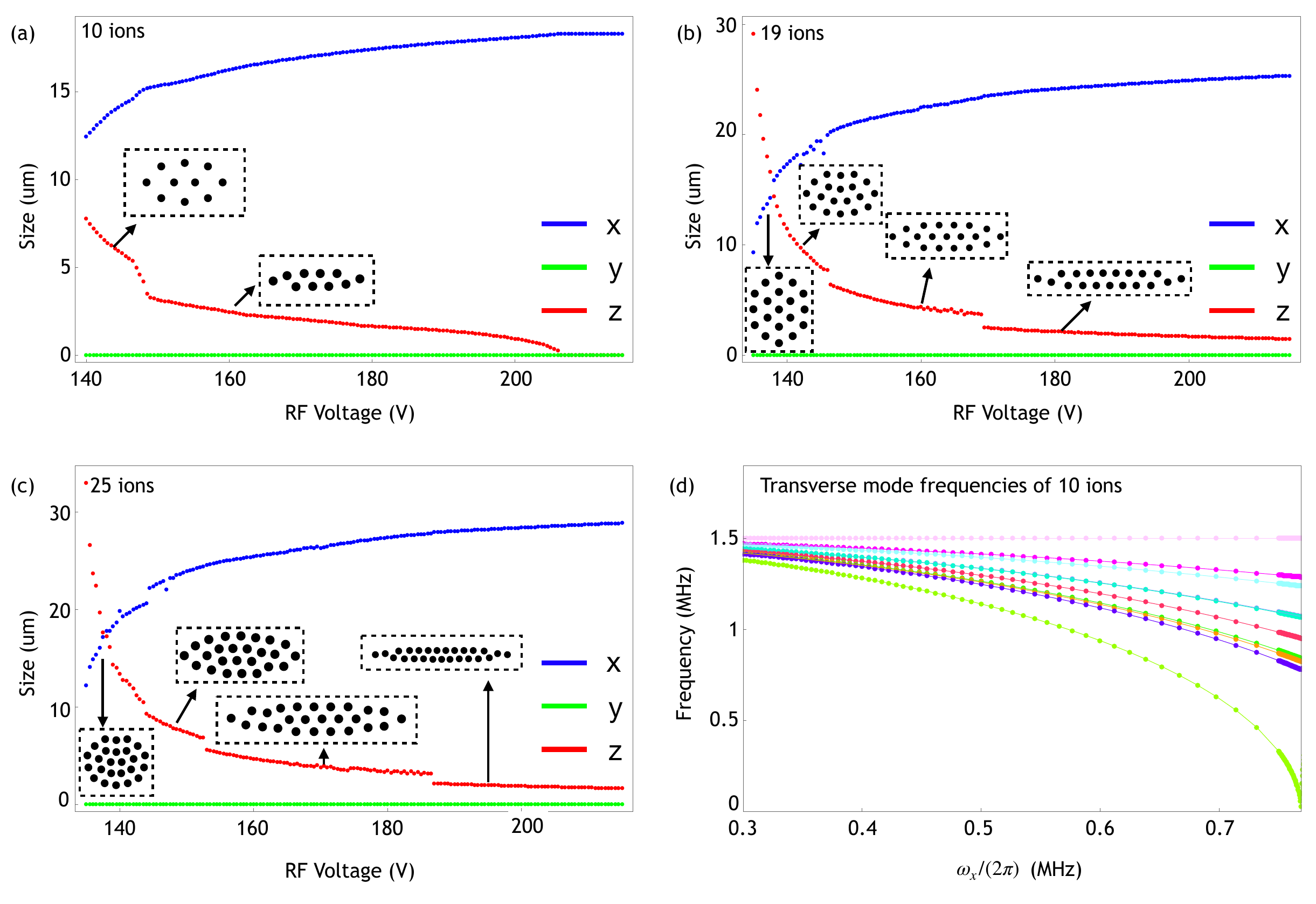}}
\caption{\label{fig:bounds}\textbf{Simulation of geometry and mode frequencies }(\textbf{a}-\textbf{c}) The relation between crystal size and the RF voltage for the cases of 10,19, and 25 ions. Here we define the size of crystal as the maximal coordinate difference in the x, y, or z axes among the ions. The zero value of the crystal size along the y-axis shows the crystal is confined in 2D on the x-z plane. And when the size of z axis becomes zero, the ions form a linear chain. The sudden jumps of the crystal size indicates a structure phase transition. (\textbf{d}) If we squeeze the crystal formed by 10 ions along the x, z-axis, defined in Fig.\ref{fig:trap} (a), by increasing $\omega_{\rm x}$ and $\omega_{\rm z}$, and keep the ratio $\omega_{\rm z}/\omega_{\rm x}=1.3$ and $\omega_{\rm y}=1.5\rm{MHz}$, the frequency of the motional modes along the y-axis will become broader. And once the minimal frequency meet zero, a phase transition from 2D to 3D happens.} 
\end{figure}

We calculate the dashed lines in Fig. \ref{fig:PrinAxis}(e) using the formula $\omega_y/(2.264N)^{1/4}$ \cite{richerme2016two}, where $\omega_y$ varies with the RF voltage. When $\omega_{\rm x}$ and $\omega_{\rm z}$ are both bigger than $\omega_y/(2.264N)^{1/4}$, the ions form a 3D crystal. When  $\omega_{\rm x}$ and $\omega_{\rm z}$ are both smaller than the bound, the ions from a 2D crystal. However, when two frequencies are not larger or smaller than the bound at the same time, there is no simple expression of the critical point for the phase transition from a 2D crystal to a 3D crystal. For example, if one of the modes is below the bounds while the another above, the ions can still form a 2D crystal. We can imagine such a situation from a homogeneous crystal where $\omega_{\rm x}=\omega_{\rm z} > \omega_{\rm y}/(2.264N)^{1/4}$. In this case the ions form a 3D crystal not 2D, but if we release the confinement along the x-axis by lowering $\omega_{\rm x}$, at a certain $\omega_{\rm x}$, the ions can form a 2D crystal. We verify this situation for 10, 19, and 25 ions by numerically simulating the equilibrium positions of the ions and study the structures of the crystals if they are in 2D as shown in Fig. \ref{fig:bounds}. 


As mentioned in the main text, in the region near the phase transition from 2D to 3D, the minimal frequency of the transverse modes will tends to zero. We also numerically study this behavior on a 10-ion 2D crystal and the show the result in Fig. \ref{fig:bounds} (d).

\onecolumngrid

\twocolumngrid
\bibliography{references}
\end{document}